\documentclass[12pt,twoside]{article}
\usepackage{graphicx}
\usepackage{amssymb}
\usepackage{amsmath}
\begin{document}

\def\cl{\centerline}
\def\bd{\begin{description}}
\def\be{\begin{enumerate}}
\def\ben{\begin{equation}}
\def\benn{\begin{equation*}}
\def\een{\end{equation}}
\def\eenn{\end{equation*}}
\def\benr{\begin{eqnarray}}
\def\eenr{\end{eqnarray}}
\def\benrr{\begin{eqnarray*}}
\def\eenrr{\end{eqnarray*}}
\def\ed{\end{description}}
\def\ee{\end{enumerate}}
\def\al{\alpha}
\def\b{\beta}
\def\bR{\bar\R}
\def\bc{\begin{center}}
\def\ec{\end{center}}
\def\d{\dot}
\def\D{\Delta}
\def\del{\delta}
\def\ep{\epsilon}
\def\g{\gamma}
\def\G{\Gamma}
\def\h{\hat}
\def\iny{\infty}
\def\La{\Longrightarrow}
\def\la{\lambda}
\def\m{\mu}
\def\n{\nu}
\def\noi{\noindent}
\def\Om{\Omega}
\def\om{\omega}
\def\p{\psi}
\def\pr{\prime}
\def\r{\ref}
\def\R{{\bf R}}
\def\ra{\rightarrow}
\def\s{\sum_{i=1}^n}
\def\si{\sigma}
\def\Si{\Sigma}
\def\t{\tau}
\def\th{\theta}
\def\Th{\Theta}

\def\vep{\varepsilon}
\def\vp{\varphi}
\def\pa{\partial}
\def\un{\underline}
\def\ov{\overline}
\def\fr{\frac}
\def\sq{\sqrt}

\def\WW{\begin{stack}{\circle \\ W}\end{stack}}
\def\ww{\begin{stack}{\circle \\ w}\end{stack}}
\def\st{\stackrel}
\def\Ra{\Rightarrow}
\def\R{{\mathbb R}}
\def\bi{\begin{itemize}}
\def\ei{\end{itemize}}
\def\i{\item}
\def\bt{\begin{tabular}}
\def\et{\end{tabular}}
\def\lf{\leftarrow}
\def\nn{\nonumber}
\def\va{\vartheta}
\def\wh{\widehat}
\def\vs{\vspace}
\def\Lam{\Lambda}
\def\sm{\setminus}
\def\ba{\begin{array}}
\def\ea{\end{array}}
\def\bd{\begin{description}}
\def\ed{\end{description}}
\def\lan{\langle}
\def\ran{\rangle}

\bc

\textbf{Entanglement Capacity of Nonlocal Hamiltonians : A Geometric Approach}

\vspace{.2 in}

\textbf{ Behzad Lari\footnote{Electronic address:behzadlari1979@yahoo.com}, Ali Saif M. Hassan\footnote{Electronic address: alisaif73@gmail.com, alisaif@physics.unipune.ernet.in} and  Pramod S. Joag\footnote{Electronic address: pramod@physics.unipune.ernet.in}}\\
 Department of Physics, University of Pune, Pune, India-411007.

\ec

\vspace{.2in}

We develop a geometric approach to quantify the capability of creating entanglement for a general physical interaction acting on two qubits. We use the entanglement measure proposed by us for $N$-qubit pure states (Phys. Rev. A \textbf{77}, 062334 (2008)). This geometric method has the distinct advantage that it gives the experimentally implementable criteria to ensure the optimal entanglement production rate without requiring a detailed knowledge of the state of the two qubit system. For the production of entanglement in practice, we need criteria for optimal entanglement production which can be checked {\it in situ} without any need to know the state, as experimentally finding out the state of a quantum system is generally a formidable task. Further, we use our method to quantify the entanglement capacity in higher level and multipartite systems. We quantify the entanglement capacity for two qutrits and find the maximal entanglement generation rate and the corresponding state for the general isotropic interaction between qutrits, using the entanglement measure of $N$-qudit pure states proposed by us (Phys. Rev. A \textbf{80}, 042302 (2009)). Next we quantify the genuine three qubit entanglement capacity for a general interaction between qubits. We obtain the maximum entanglement generation rate and the corresponding three qubit state for a general isotropic interaction between qubits. The state maximizing the entanglement generation rate is of the GHZ class. To the best of our knowledge, the entanglement capacities for two qutrit and three qubit systems have not been reported earlier. \\


\vspace{.2 in}

\bc
\textbf{I. INTRODUCTION}\\
\ec
It is by now well established that entanglement in multipartite quantum systems is a physical resource used to perform a variety of information processing tasks [1] as well as novel communication protocols [2]. A quantum system evolves to generate entanglement provided its parts interact. For such an interaction, the Hamiltonian of the total system is not just a sum of the Hamiltonians pertaining to each part (local Hamiltonians).   Thus, for a bipartite system $AB,$ $H_{AB}\neq H_{A}+H_{B}$ but has a term which couples the two parts $A$ and $B.$
Together with local operations, this coupling can be used to generate entanglement [3,4,5], to transmit classical and quantum information [4,6,7,8] and more generally, to simulate the dynamics of some other Hamiltonian (say $H^{\prime}_{AB}$) and thus to perform arbitrary unitary gates on the composite space $\mathcal{H}_{AB}=\mathcal{H}_{A}\otimes \mathcal{H}_{B}$ [9,10,11].

A lot of experimental work is devoted to creating entangled states of quantum systems, including those in quantum optics, nuclear magnetic resonance and condensed matter physics [12]. Determining the ability of a system to create entangled states provides a benchmark of the ``quantumness'' of the system. Furthermore, such states can ultimately be put to some information processing task like superdense coding [13], or quantum teleportation [14].

The theory of optimal entanglement generation can be approached in different ways. Ref. [3] considers {\it single shot} capacities. For two qubit interaction, without any ancilla qubits, Ref.[3] presents a closed form expression for the entangling capability and optimal protocols by which it can be achieved. In contrast, Ref.[4] considers the {\it asymptotic} entanglement capacity, allowing the use of ancillary systems and shows that when ancillas are allowed, the single shot and asymptotic capacities are in fact the same. However, such capacities are difficult to calculate because the ancillary systems may be arbitrarily large. In this paper we exclusively deal with the single shot entanglement capacity. Throughout this paper, we take $\hbar=1.$ 

The paper is organized as follows. In section II we deal with entanglement capacity for two qubit states while in section III we deal with this problem involving two qutrits. In section IV we address the problem of the entanglement capacity involving the genuine tripartite entanglement for three qubits. The discussion of the results for two qubit, two qutrit and three qubit cases is included separately in sections II, III and IV respectively. 

\bc
\textbf{II. THE TWO QUBIT CASE}\\
\ec

We develop a geometric approach to calculate the entanglement capacity of any two qubit system interacting via a Hamiltonian which is locally equivalent to $$H_{I}= \mu_1\si^{A}_1\otimes \si^{B}_1+ \mu_2\si^{A}_2\otimes \si^{B}_2+ \mu_3\si^{A}_3\otimes \si^{B}_3.\eqno{(1)}$$ Here $\mu_1\geq\mu_2\geq\mu_3.$ In this section we omit qubit and system identifiers $A,B$ and $AB.$

We define the single shot entanglement capacity by
\begin{displaymath}
\Gamma^{max}=\max_{\substack{|\psi\ran\in\mathcal{H}_{I}}}
\lim_{t\to 0}\frac{E\Big(e^{-iHt}|\psi\ran\Big)-E (|\psi\ran)}{t}.\eqno{(2)}
\end{displaymath}
The Hamiltonian $H_{I}$ in Eq. (2) is given by Eq. (1).
$E(|\psi\ran)$ in Eq. (2)  stands for the two qubit pure state entanglement measure given by us and is shown to have all the essential (as well as many desirable, e.g., superadditivity and continuity) properties expected of a good entanglement measure [15]. For a $N$-qubit pure state $|\psi\ran ,$
$$E(|\psi\ran)=||\mathcal{T}^{(N)}||-1$$ where $||\mathcal{T}^{(N)}||$ is the Hilbert-Schmidt (Euclidean) norm of the $N$ way array $\mathcal{T}^{(N)}$ occurring in the Bloch representation of $|\psi\ran\lan\psi|$ [15,16].

The scenario we address, is as follows [3].  The idea is to supplement the interaction Hamiltonian $H_{I}$ with appropriate local unitary operations in such a way that the state of the qubits at any time $t$ is precisely $|\psi_{E(t)}\ran$, for which the increase of entanglement is optimal. In order to construct such a procedure, we consider the evolution given by $H_{I}$ to proceed in very small time steps $\delta t$. Let us also assume that the qubits are initially disentangled. Using local operations,  we can always prepare the state $|\psi_0\ran$ that is, the product state which most efficiently becomes entangled under the action of $H_{I}$. After a time step $\delta t$, the state will change and its entanglement will increase to $\delta E$. Then, we use (fast) local unitary operations to transform the new state of the qubits into the state $|\psi_{\delta E}\ran$ for which $\Gamma$ is optimal. Note that this is always possible, since for qubits all states with the same value of $E$, say $\delta E$, are connected by local unitary transformations. By proceeding in the same way after every time step, and taking the continuous time limit $\delta t\rightarrow 0$, we obtain that the state of the qubits at time $t$ is always the optimal one, $|\psi_{E(t)}\ran$. Obviously, in an experimental realization, this procedure requires that we can apply the appropriate local transformations in times which are short compared to the typical time scale $\tau_{H_{I}}$ associated with $H_{I}$, $\tau_{H}=(e_{max}-e_{min})^{-1}$, where $e_{max}$ and $e_{min}$ are the maximum and minimum eigenvalues of $H_{I}$. Note that Eq. (2) defines the entanglement capacity as the maximum achievable entanglement rate for a given system with given interactions. We are also interested in finding the state $|\psi_{max}\ran$ for which the entanglement rate is maximum, (denoted by $\Gamma^{max}$ in Eq. (2)).

We consider two qubits interacting via the Hamiltonian $H_{I}$ in Eq. (1), which represents general interaction between two qubits [3]. First we find the entanglement rate $\Gamma$ given by
\begin{displaymath}
\Gamma = \lim_{t\to 0}\left[\frac{E\Big(e^{-iHt}|\psi\ran\Big)-E (|\psi\ran)}{t}\right] \equiv \frac{dE}{dt}.\eqno{(3)}
\end{displaymath}
Here $|\psi\ran$ is given by a general two qubit state in the Bloch representation [15,16],
$$\rho=|\psi\ran\lan\psi|=\frac{1}{4}\left(I\otimes I + \sum_{k}r_{k} \si_{k}\otimes I + \sum_{l}s_{l} I\otimes \si_{l}
+ \sum_{k,l}\tau_{kl}\si_{k}\otimes \si_{l}\right),\eqno{(4)}$$
where $\si_{k,l},\;k,l=1,2,3$ are the Pauli operators.
We denote by $\mathcal{T}=[\tau_{ij}]$ the correlation matrix occurring in the last term of Eq. (4). $\tau_{ij}$ are defined by $$\tau_{ij}=Tr(\si_i\otimes\si_j \rho)=\lan\psi|\si_i\otimes\si_j|\psi\ran .\eqno{(5)}$$ $r_{k}$ and $s_{l}$ are the components of the Bloch vectors [16] of the reduced density operators $\rho_{A}$ and $\rho_{B}$ respectively, given by
$$r_{k}=Tr(\si_{k}\rho_{A})=\lan\psi|\si_k \otimes I |\psi\ran ,\eqno{(6a)}$$ $$s_{l}=Tr(\si_{l}\rho_{B})=\lan\psi| I \otimes \si_{l} |\psi\ran .\eqno{(6b)}$$ We define the entanglement of the state $|\psi\ran$ as [15] $$E(|\psi\ran)=||\mathcal{T}||-1,\eqno{(7)}$$ where $||\mathcal{T}||=\sqrt{\sum_{ij=1}^{3}\tau_{ij}^2}$ is the Euclidean norm of $\mathcal{T}.$ For two qubits, this measure is related to concurrence [15] and hence to the Von Neumann entropy of the reduced density matrix.

After finding $\Gamma$ we maximize it, using a simple geometric argument. It is heartening to see that the scenario described above emerges naturally out of this geometric method.

The entanglement rate $\Gamma$ is given by (See Eq. (3)) $$\Gamma= \frac{dE}{dt}=\frac{d||\mathcal{T}||}{dt}=
\frac{1}{||\mathcal{T}||} \sum_{ij}\tau_{ij}\dot{\tau}_{ij},$$ with $\tau_{ij}$ given by Eq. (5). We evaluate $\dot{\tau}_{ij}$ as follows. $$\dot{\tau}_{ij}=\frac{d\tau_{ij}}{dt}=\frac{d}{dt}(Tr(\si_i\otimes\si_j \rho))=Tr(\si_i\otimes\si_j\fr{d\rho}{dt}).$$ We now use the equation of motion ,
$$i\frac{d\rho}{dt}=[H_{I},\rho],$$ where the Hamiltonian $H_{I}$ is defined via Eq. (1), to get [17],
 $$\fr{d\t_{ij}}{dt}=-iTr(\si_i\otimes\si_j [H_{I},\rho])=iTr(H_{I}[\si_i\otimes\si_j,\rho]).$$  Substituting $\rho$ from Eq. (4) and using the commutation relations [17] $$[\si_i\otimes\si_j,\si_k\otimes\si_l]=\frac{1}{2}[\si_i,\si_k]\otimes\{\si_j,\si_l\}
+\frac{1}{2}\{\si_i,\si_k\}\otimes [\si_j,\si_l],$$ we get, using $[\si_i,\si_j]=2i\vep_{ijk}\si_k ,$ $\{\si_i,\si_j\}=
2\delta_{ij}$ and the expression of $H_{I}$ in Eq. (1), after a bit of algebra, $$\frac{d\tau_{ij}}{dt}=-2\left[\sum_{k,n}
r_k\vep_{ikn}\mu_n\del_{nj}+ \sum_{l,n} s_l\vep_{jln}\mu_n\del_{ni}\right].$$ This gives $$\sum_{ij}\tau_{ij}\dot{\tau}_{ij}=-2\left[\sum_{i,k,n}\tau_{in}r_k\vep_{ikn}\mu_n+ \sum_{j,k,n}\tau_{nj}s_l\vep_{jln}\mu_n\right].$$ Thus we get, for the entanglement rate $\G ,$
$$\G = \frac{2}{||\mathcal{T}||}\sum_{n} \left[(\vec{r}\times\vec{\tau}_{:n})_{n} + (\vec{s}\times\vec{\tau}_{n:})_{n} \right]\mu_n .\eqno{(8)}$$ Here $\vec{\tau}_{:n}$ and $\vec{\tau}_{n:}$ are, respectively, the $n$ th column and row vectors of the correlation matrix $\mathcal{T}=[\tau_{ij}] .$

The entanglement generation rate $\G$ expressed in Eq. (8) is obtained via the temporal evolution of the initial state by the interaction Hamiltonian $H_{I}.$
This expression for $\G$ does not depend on any local unitary transformation applied to a qubit. Following the general scenario described above, (see the third paragraph of this section), we now lock on to an instant of time and apply the local unitary transformations to qubits, in order to find the conditions for optimal $\G$ and the corresponding two qubit state $|\psi_{E}\ran.$ The experimental meaning of this sentence is described as a part of the scenario above. In the geometrical approach we have adopted, local unitary transformations amount to rotations of vectors in Eq. (8), which are the vectors in the Bloch space of individual qubits. We expect the entanglement to remain unultered by the local unitaries, which turns out to be the case. The entanglement measure in Eq.(7) is not affected by local unitaries, as proved in [15].

Obviously, $\G$ will be maximum if the components of the vector products occurring in Eq. (8) are replaced by the magnitudes of these vector products and the factors in these products are mutually perpendicular. Geometrically, this means that the vector products themselves are in the directions of the components occurring in Eq. (8) with the other two orthogonal components zero. For example, $(\vec{r}\times\vec{\tau}_{:1})$ is along its first component, i.e. along $x$ axis, with its $y$ and $z$ components zero. Thus, in order to maximize the first term in Eq. (8), namely,
$$\sum_{n} (\vec{r}\times\vec{\tau}_{:n})_{n}\mu_n=(\vec{r}\times\vec{\tau}_{:1})_{1}\mu_1+(\vec{r}\times\vec{\tau}_{:2})_{2}\mu_2+(\vec{r}\times\vec{\tau}_{:3})_{3}\mu_3,$$ we must have vectors $(\vec{r}\times\vec{\tau}_{:1}),$
$(\vec{r}\times\vec{\tau}_{:2})$ and $(\vec{r}\times\vec{\tau}_{:3})$ along $x,y,z$ axes respectively. This can be done only when one of the vector products is zero. Since $\mu_1\geq\mu_2\geq\mu_3,$
we choose $(\vec{r}\times\vec{\tau}_{:3})=0.$ Given the vector $(\vec{r}\times\vec{\tau}_{:1})$ along the $x$ axis and the vector $(\vec{r}\times\vec{\tau}_{:2})$ along the $y$ axis, we can choose $\vec{r}$ to be along the $z$ axis and vectors $\vec{\tau}_{:1}$ and $\vec{\tau}_{:2}$ along the $y$ and $x$ axes respectively. In exactly the same way, maximization of the second term in Eq. (8), $\sum_{n}(\vec{s}\times\vec{\tau}_{n:})_{n}\mu_n,$ makes the vector $\vec{s}$ along the $z$ axis and vectors $\vec{\tau}_{1:}$ and $\vec{\tau}_{2:}$ along the $y$ and $x$ axes respectively. Writing explicitly the components of the vector products in the expression for $\G$ (Eq. (8)) and putting $r_{1,2}=0=s_{1,2}$ we get,
$$\G=\frac{2}{||\mathcal{T}||}((-r_3\tau_{21}-s_3\tau_{12})\mu_1+(r_3\tau_{12}+s_3\tau_{21})\mu_2).$$ Since we are dealing with the two qubit pure states we have $||\vec{r}||=||\vec{s}||$ [18], so that $r_3=\pm s_3.$ Choosing
$$r_3=- s_3\eqno{(9)}$$ we get, $$\G = \frac{2}{||\mathcal{T}||}r_3(\tau_{12}-\tau_{21}) (\mu_1+\mu_2).$$
  The expression $(\tau_{12}-\tau_{21})$ becomes maximum when $$\tau_{12}=-\tau_{21}.\eqno{(10)}$$ Finally, we note that this maximization procedure does not change $||\mathcal{T}||=\sqrt{\sum_{n}||\tau_{:n}||^2}$ and hence the entanglement value given by Eq.(7). Further, choosing the cross products along their components appearing in Eq. (8) corresponds to the rotations in Bloch space, generating local unitaries on the system. Therefore, the maximum of $\G$ over the states with same entanglement, that is, $\G_{E},$ is given by $$\G_{E}=\frac{4}{||\mathcal{T}||}r_3\tau_{12}(\mu_1+\mu_2).\eqno{(11)}$$

To get the state $|\p_{E}\ran$ corresponding to $\G_{E},$ we seek the state satisfying conditions Eq. (9) and Eq. (10). We start with the general state $|\p\ran =\sum_{i,j=0}^{1}c_{ij}|ij\ran$ and calculate $\tau_{12}$ and $\tau_{21}.$ In order to satisfy Eq.(10), the state $|\p\ran$ should be $$|\p_{E}\ran = |c_{01}| |01\ran + i|c_{10}| |10\ran\;;\;
|c_{01}|^2+|c_{10}|^2=1,\eqno{(12)}$$ which is the same as $|\p_{E}\ran$ obtained in Ref [3] if we identify $|c_{01}|=\sqrt{p}.$ Further, we can write $\G_{E}$ (Eq. (11)) as the product of two factors $$\G_{E}=f(p)h_{max}$$ with $$h_{max}=(\mu_1+\mu_2)$$ and $$f(p)=\frac{4 r_3\tau_{12}}{||\mathcal{T}||}.\eqno{(13)}$$ To get $f(p)$ as a function of $p,$ we calculate $r_3,$ $\tau_{12}$ and $||\mathcal{T}||$ using the state $|\p_{E}\ran$ (Eq. (12)) so that
$$f(p)=\frac{4 r_3\tau_{12}}{||\mathcal{T}||}=\frac{8(1-2p)\sqrt{p(1-p)}}{\sqrt{1+8p(1-p)}}.$$ 
Fig.(1a) depicts this $f(p)$ verses $p,$ while Fig.(1b) plots the analogous $f(p)$ obtained using Von Neumann entropy as the entanglement measure. Note that $f(p)$ and hence $\G_{E}$
vanishes for the maximally entangled state ($p=\frac{1}{2}$) about which it is antisymmetric $f(\frac{1}{2}+x)=-f(\frac{1}{2}-x).$ We see that, as $p$ increases from $0$ to $\fr{1}{2}$, $\G_E>0$ makes the entanglement increase, until is maximal at $p=\fr{1}{2}$, after which $\G<0$, making entanglement decrease to zero as $p$ approaches $1$.

\begin{figure}[!ht]
\begin{center}
\includegraphics[width=11cm,height=8.3cm]{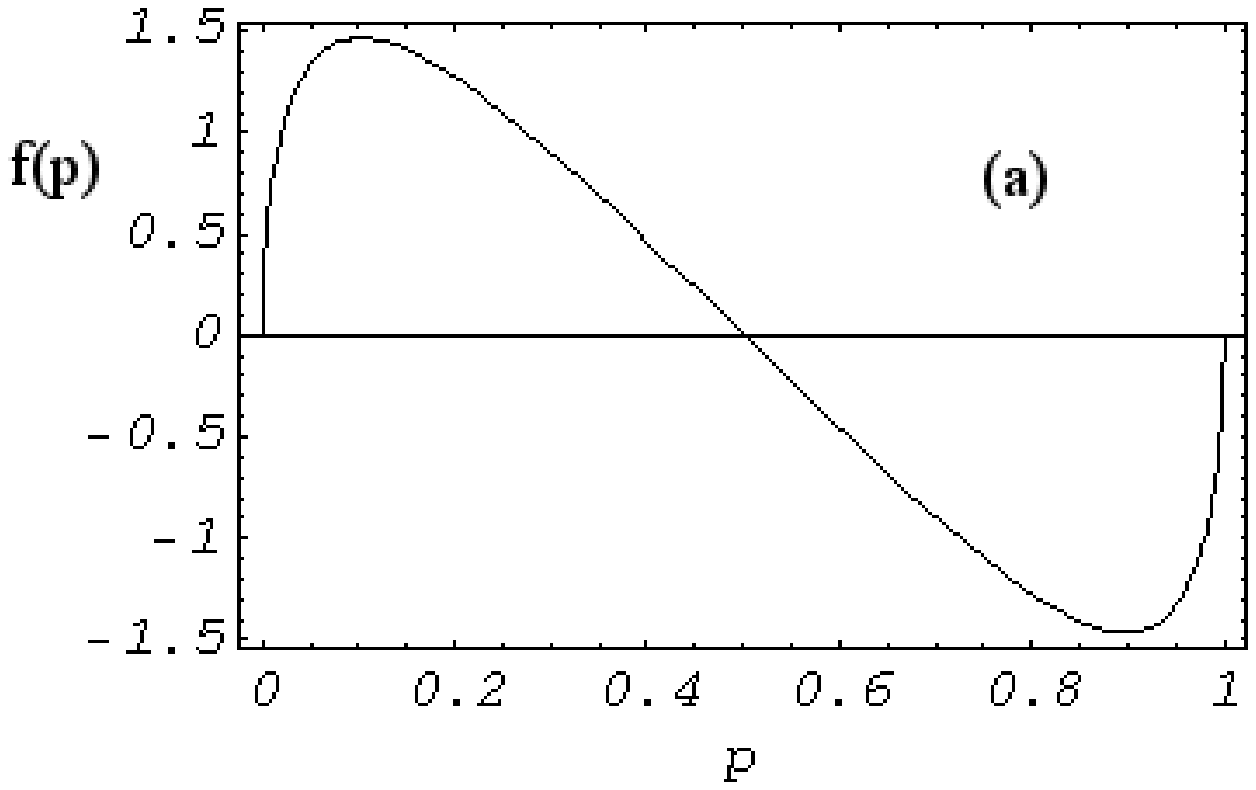}

Fig. 1a
\end{center}
\end{figure}

\begin{figure}[!ht]
\begin{center}
\includegraphics[width=10.5cm,height=8.3cm]{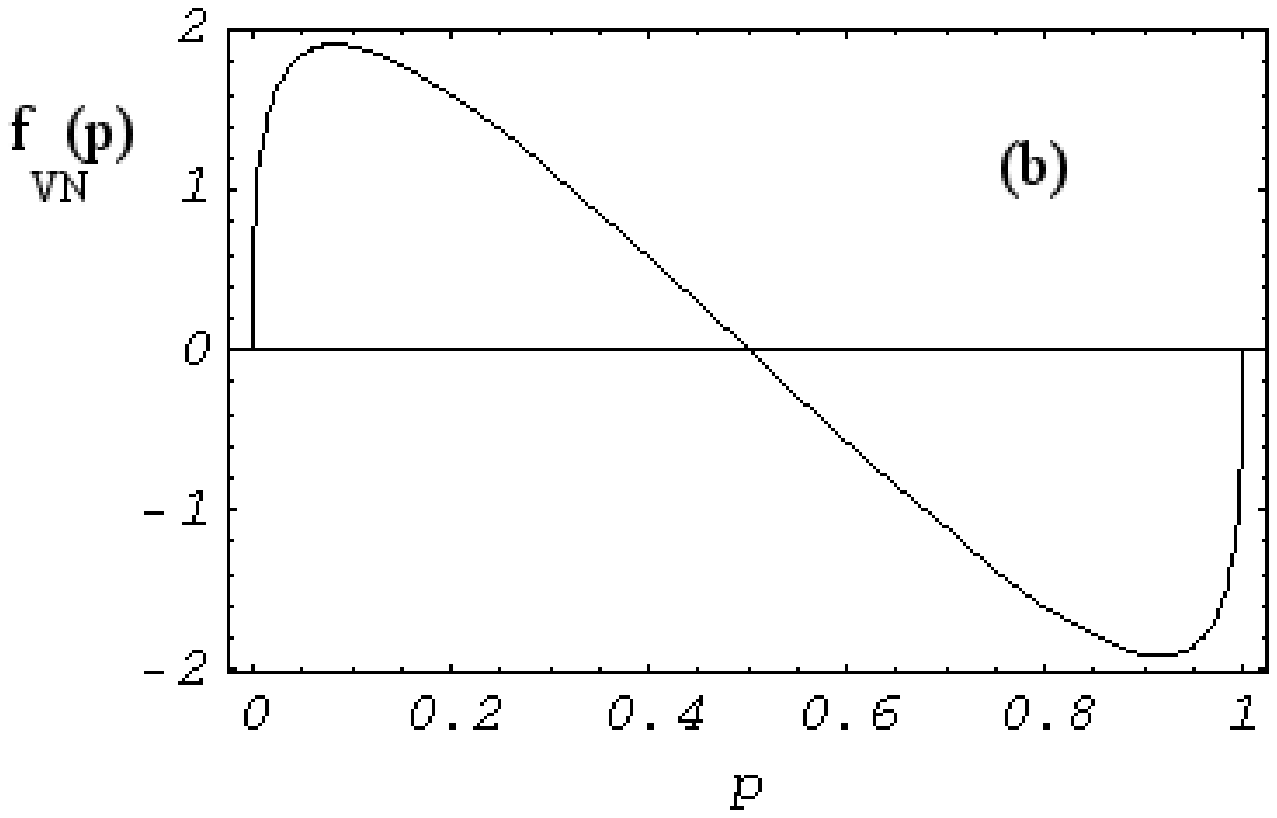}

Fig. 1b

Fig.1: (a) $f(p)$ vs. $p$ for entanglement measure in Eq. (7) and (b) $f_{VN}(p)$ vs. $p$ for Von Neumann entropy of the reduced density matrix (see text).
\end{center}
\end{figure}

Thus we see that, $\G_{E}$ is the product of the function which depends only on the state, (via $p$) and the factor $h_{max}$ which depends only on the interaction strengths $\mu_1$ and $\mu_2$, that is, on the interaction Hamiltonian. Note that $h_{max}$ is independent of the entanglement measure. The form of $f(p)$ for the entanglement measure in Eq.(7) and that for the Von Neumann entropy, (Fig.(1b)) also turns out to be the same. To get  $\Gamma^{max}$ we have to find $p_0$ at which $f(p)$ is maximum. To do this, we invoke the relation between $f_{VN}(p),$ which is the analog of $f(p)$ in Eq. (13) obtained via Von Neumann entropy of the reduced density operator of the two qubit pure state [3] and $f(p)$ in Eq. (13) obtained via $E(|\psi\ran)$ in Eq. (7). This is 
\setcounter{equation}{13}
\begin{equation}
f_{VN}(p)=f(p)\left(\frac{dE_{VN}}{dp}\Big/\frac{dE}{dp}\right), \label{eqn14}
\end{equation}
where $E_{VN}$ is the Von Neumann entropy of the reduced density operator and $E$ is given by Eq. (7). Maximizing the RHS of Eq. (\ref{eqn14}) we get 
$p_0\approx 0.0832217$ and $\G^{max}\approx 1.9123$. The state $|\p_{max}\ran$ corresponding to $\Gamma^{max}$ is the state $|\p_{E}\ran$ with $p=p_0.$

From the definitions of $\tau_{ij}$ and $r_k$(Eq. (5) and (6a)), these quantities are averages of the Pauli operators in a state $|\psi\ran$, which can be obtained using experimentally measured values of the corresponding operators on two qubits. Therefore, {\it the geometric method presented here has the advantage that the conditions for $\G_{E},$ Eqs. (9) and (10), can be tested experimentally giving us an experimental way to check out whether the system has reached the state $|\p_{E}\ran$.} The value of $\G_{E}$ can also be experimentally estimated via Eq. (11), for given $\mu_1+\mu_2.$  Further, the function $f(p)$ can be estimated experimentally via Eq. (13) as the system evolves, under the given Hamiltonian, toward $|\p_{max}\ran,$ or under the local unitaries toward $|\p_{E}\ran$. {\it These experimental estimations can be carried out without a detailed a priori knowledge of the quantum state at any time during the evolution of the two qubit system.} These facts can be of great advantage in a practical implementation of any scheme to entangle two qubits interacting via some Hamiltonian or quantum gates [19]. For the production of entanglement in practice, we need criteria for optimal entanglement production which can be checked {\it in situ} without any need to know the state, as experimentally finding out the state of a quantum system is generally a formidable task. We note that, in order to achieve such an experimental determination of optimal entanglement production rate using the model in ref [3], we have to experimentally obtain the values of the Schmidt coefficients of the evolving two qubit state, which requires the experimental determination of the two qubit state itself. This requires more experimental effort and resources ($d+1$ different joint measurements, $d=$ dimension of the joint Hilbert space [20]) as compared to measuring the quantities in Eqs. (9) and (10), which are simply the average values of the Pauli operators in the state.  \\

\bc
\textbf{III. THE TWO QUTRIT CASE}\\
\ec

The entanglement measure in Eq. (7) can be generalized to the $N$ qudit pure states which satisfies all the essential (and many desirable, e.g., superadditivity and continuity) properties expected of a good entanglement measure [21]. Therefore, we can use it to obtain the entanglement generation rates for the multipartite $d$ level systems. Here we find the entanglement generation rate for two qutrits (labeled $A$ and $B$) interacting via the Hamiltonian
\begin{equation}
H_{I}= \sum_{p=1}^{8}\mu_p\la^{A}_p\otimes \la^{B}_p, \label{eqn29} 
\end{equation}
where $\mu_p$s are the interaction strengths satisfying $\mu_k\geq\mu_l$ for $k<l, k,l=1,\ldots, 8.$ Here $\la_p,\;\;p=1,\ldots,8$ are the generators of the $SU(3)$ group satisfying $Tr(\la_i\la_j)=2\del_{ij}$ and are characterized by the structure constants of the corresponding Lie algebra, $f_{ijk}\;\textrm{and}\;g_{ijk},$ which are, respectively, completely antisymmetric and completely symmetric.
\begin{equation}
\la_i\la_j=\fr{2}{3}\del_{ij}I_3+if_{ijk}\la_k+g_{ijk}\la_k. \label{eqn30}
\end{equation}  
Other useful relations are
\ben
4if_{jkl}=Tr([\la_i,\la_k]\la_l) \label{eq30}
\een
\ben
4g_{ikp}=Tr(\{\la_i,\la_k\}\la_p).  \label{eq31}
\een

We give here the  generators of $SU(3)$ in the $|1\ran,\;|2\ran,\;|3\ran$ basis [17] to be used below.\\
$\lambda_1=|1\ran\lan 2|+|2\ran \lan 1|$\\
$\lambda_2=-i(|1\ran\lan 2|-|2\ran \lan 1|)$\\
$\lambda_3=|1\ran\lan 1|-|2\ran \lan 2|$\\
$\lambda_4=|1\ran\lan 3|+|3\ran \lan 1|$\\
$\lambda_5=-i(|1\ran\lan 3|-|3\ran \lan 1|)$\\
$\lambda_6=|2\ran\lan 3|+|3\ran \lan 2|$\\
$\lambda_7=-i(|2\ran\lan 3|-|3\ran \lan 2|)$\\
$\lambda_8=\fr{1}{\sqrt{3}}(|1\ran\lan 1|+|2\ran \lan 2|-2 |3\ran\lan3|).$\\

The action of these generators on the basis states $\{|1\ran,|2\ran,|3\ran\}$ is given by the following.

$\lambda_1 |1\ran=|2\ran,\; \lambda_1 |2\ran=|1\ran,\; \lambda_1 |3\ran=0$\\
$\lambda_2 |1\ran=i|2\ran,\; \lambda_2 |2\ran=-i|1\ran,\; \lambda_2 |3\ran=0$\\
$\lambda_3 |1\ran=|1\ran,\; \lambda_3 |2\ran=-|2\ran,\; \lambda_3 |3\ran=0$\\
$\lambda_4 |1\ran=|3\ran,\; \lambda_4 |2\ran=0,\; \lambda_4 |3\ran=|1\ran$\\
$\lambda_5 |1\ran=i|3\ran,\; \lambda_1 |2\ran=0,\; \lambda_1 |3\ran=-i|1\ran$\\
$\lambda_6 |1\ran=0,\; \lambda_6 |2\ran=|3\ran,\; \lambda_6 |3\ran=|2\ran$\\
$\lambda_7 |1\ran=0,\; \lambda_7 |2\ran=i|3\ran,\; \lambda_7 |3\ran=-i|2\ran$\\
$\lambda_8 |1\ran=\fr{1}{\sqrt{3}}|1\ran,\; \lambda_8 |2\ran=\fr{1}{\sqrt{3}}|2\ran,\; \lambda_8 |3\ran=-\fr{2}{\sqrt{3}}|3\ran .$\\
We use these equations to get the vectors $\vec{\Lambda}^{A}$ and $\vec{\Lambda}^{B}$ in $\mathbb{R}^8$ whose components are the averages $\Lambda_{i}^{A}=\lan\psi|\la_{i}\otimes I|\psi\ran=Tr(\la_{i}\otimes I\rho),\;i=1,\ldots,8$ and $\Lambda_{i}^{B}=\lan\psi|I\otimes\la_{i}|\psi\ran=Tr(I\otimes\la_{i}\rho),\;i=1,\ldots,8$ respectively,
where $|\psi\ran,$ ($\rho=|\psi\ran\lan\psi|$) is a two qutrit pure state (see Eq.(\ref{eqn36}) and the discussion following it).

The pure state entanglement for two qutrits is given by [21],
\begin{equation}
E(|\psi\ran)=||\mathcal{T}||-3  \nn
\end{equation}
where $||\mathcal{T}||=\sqrt{\sum_{ij=1}^{8}\tau_{ij}^2}$ is the Euclidean norm of $\mathcal{T}.$ The general two qutrit pure state $\rho$ has the following Bloch representation.
\benr
\rho &=& |\psi\ran\lan\psi|    \nn \\
 &=&\frac{1}{9}\left(I^{A}\otimes I^{B} +\fr{3}{2}\left( \sum_{k}\lan \la_{k}^{A}\ran \la_{k}^{A}\otimes I^{B} + \sum_{l}\lan \la_{l}^{B}\ran I^{A}\otimes \la_{l}^{B}\right) + \fr{9}{4}\sum_{k,l}\tau_{kl}\la_{k}^{A}\otimes \la_{l}^{B}\right),  \nn
\eenr
Here $\lan\la^{A,B}\ran=Tr(\rho_{A,B}\la^{A,B})$ with $\la^{A,B}$ and $\rho_{A,B}$ (the reduced density operator) apply to the qutrit A and B respectively, while $\tau_{kl}=(9/4)Tr(\la_k^{A}\otimes\la_l^{B}\rho).$ The definitions of $\dot{\tau}_{kl}$ and $\G$ are 
\begin{equation}
\G=\frac{1}{||\mathcal{T}||} \sum_{ij}\tau_{ij}\dot{\tau}_{ij},  \nn
\end{equation}
\begin{equation}
\dot{\tau}_{ij}=iTr(H[\la_i\otimes\la_j, \rho]),  \nn
\end{equation}
where we have used the Heisenberg equation of motion as in the two qubit case. Using Eq. (\ref{eqn30}), (\ref{eq30}), (\ref{eq31}) and the elements of the tensors $f_{ijk}$ and $g_{ijk}$ in [17], we get, after some algebra, the following expression for $\G$
\ben
\G=-3\left(\fr{1}{||\mathcal{T}||}\right)\sum_{k,p,l=1}^{8}\mu_p f_{klp}\left(\tau_{kp}\la_l^{A}+\tau_{pk}\la_l^{B}\right). 
\label{eqn35}
\een
Expanding the sum in Eq. (\ref{eqn35}) and rearranging, we get,
\ben
\G= -3\left(\fr{1}{||\mathcal{T}||}\right)\sum_{S}\al(S)\sum_{p\in S}\mu_p[(\vec{\tau}_{:p}\times\vec{\la}^{A})_p+
(\vec{\tau}_{p:}\times\vec{\la}^{B})_p], \label{eqn36}
\een
where $S$ runs over the triplets 
$$(1,4,7),(2,1,6),(3,1,5),(3,2,4),(2,5,7),(3,7,6),(5,4,6),(3,6,8),(2,5,8)$$ and $\al(S)$ has
values $1,1/2,1/2,1/2,1/2,1/2,1/2,\sq{3}/2,\sq{3}/2$ respectively for these triplets. $\vec{\tau_{:p}}$ and $\vec{\tau_{p:}}$ 
are the vectors in $\mathbb{R}^3$ with $p\in S$ where $S$ is one of the above triplets and the index $:$ varies over a given $S$ for fixed $p.$ $\vec{\la}^{A,B}$  are vectors in $\mathbb{R}^3$ respectively comprising the components of $\vec{\Lambda}^{A,B}$ indexed by one of the triplets $S.$ There are in all $54$ terms in Eq. (\ref{eqn36}). Unfortunately, all these terms are coupled and a simple geometrical procedure to maximize $\G,$ as in the two qubit case, seems very difficult. However, it is straightforward to maximize $\G$ numerically over the coefficients $c_{ij}, \;\;i,j=0,1,2$, by expressing all the terms in the expression for $\G$ (Eq. (\ref{eqn36})) as averages in the general two qutrit state $|\psi\ran=\sum_{ij}c_{ij}|ij\ran,\;\;i,j=0,1,2.$ We can carry out the numerical maximization for the general Hamiltonian in Eq. (\ref{eqn29}), where the strengths of interaction $\mu_k$ have different values. In that case, $\G$ does not have the simple structure analogous to $\G=f(p)h_{max}$ as in the two qubit case. Therefore, we assume isotropic interactions so that all interaction strengths are equal to a common value $\mu .$ In this case, $\G$ has a simple form 
\ben
\G=h_{max}(c_{ij};\;i,j=0,1,2)\mu.\nn
\een
Therefore, we maximize $\G$ assuming the isotropic interactions. The result is
\ben
\G^{max}\approx 3.90495\mu  \nn
\een
 and the corresponding (normalized) state is given by $$c_{00}=-0.28317+i0.148948;\;c_{01}=-0.433055+i0.382479;\;c_{02}=-0.117778+i0.274948$$ $$c_{10}=0.0625717-i0.144584;\;c_{11}=0.102783-i0.0787094;\;c_{12}=-0.340939-i0.324717 $$ $$c_{20}=0.25066-i0.167261;\;c_{21}=0.0344755-i0.244282;\;c_{22}=0.227159-i0.088347.$$
After converting this state to the Schmidt canonical form we get the state giving the maximal entanglement generating rate for two qutrits, under a general isotropic interaction, as 
\ben
|\psi_{max}\ran=0.884297|00\ran+0.448838|11\ran+0.128697|22\ran. \nn
\een 
We find that $E(|\psi_{max}\ran)=0.677882.$ This shows that, in order to increase the entanglement of a two qutrit system in an optimal way, it is better to start with an initially entangled state rather than a product state, at least when all the interaction strengths in $H_{I}$ (Eq.(\ref{eqn29}))
are equal. We also note that the optimal entanglement $E(|\psi_{max}\ran)$ is independent of $H_{I},$ provided, again, that all interaction strengths in $H_{I}$ (Eq.(\ref{eqn29})) are equal. \\    

\bc
\textbf{IV. THE THREE QUBIT CASE}\\
\ec

We now deal with the problem of entanglement generation capacity for three qubits. We emphasize that this is the capacity to generate genuine three qubit entanglement and not the bipartite entanglement between any two parts of the three qubit system. In this case also, the entanglement measure given by Eq. (7) can be used as this entanglement measure applies to $N$-qubit pure states and has all the essential (and many desirable, eg superadditivity and continuity) properties expected of a good entanglement measure [15]. For the three qubit case, $\tau$ in Eq. (7) is the three qubit correlation tensor appearing in the Bloch representation of the state. Here $\tau$ is a three way array while for two qubits $\tau$ was a matrix. The Bloch representation of a general three qubit pure state is  
{\setlength\arraycolsep{2pt}
\begin{eqnarray}
\rho & = & |\psi\ran\lan\psi|=\frac{1}{8}\big(I\otimes I\otimes I + \sum_{l}r_l  \si_l\otimes I\otimes I + \sum_{n}s_n  I\otimes \si_n\otimes I 
+ \sum_{m}q_m  I\otimes I\otimes \si_m + {}  \nn \\
& & {} +\sum_{ln}t_{ln}^{(AB)}\si_l\otimes \si_n \otimes I + \sum_{lm}t_{lm}^{(AC)}\si_l\otimes I\otimes \si_m
 +\sum_{nm}t_{nm}^{(BC)}I\otimes \si_n\otimes \si_m + {} \nn \\
 & & {} +\sum_{lmn}\tau_{lmn}\si_l\otimes\si_n\otimes \si_m \big).  \label{eqn15}
\end{eqnarray}}
Here $\tau=[\tau_{ijk}]$ is a three way array while $t^{(\cdot\cdot)}=[t^{(\cdot\cdot)}_{ij}]$ are matrices. The definitions of various symbols in $\rho$ are as follows. 
\begin{eqnarray}
r_l & = & Tr(\si_{l}^{A}\rho_{A})  =  Tr(\si_l^{A}\otimes I\otimes I \;\rho)\nonumber \\
s_n & = & Tr(\si_{n}^{B}\rho_{B})  =  Tr( I\otimes \si_n^{B}\otimes I \;\rho)\nonumber \\
q_m & = & Tr(\si_{m}^{C}\rho_{C})  = Tr(I\otimes I\otimes \si_m^{C}   \;\rho),  \nn
\end{eqnarray}
\begin{eqnarray}
t_{ln}^{AB} & = & Tr(\si_l^{A}\otimes\si_{n}^{B}\rho_{AB}) = Tr(\si_l^{A}\otimes\si_{n}^{B}\otimes I^{C}\rho) \nn \\
t_{lm}^{AC} & = & Tr(\si_l^{A}\otimes\si_{m}^{C}\rho_{AC}) = Tr(\si_l^{A}\otimes I^{B} \otimes\si_{m}^{C}\rho) \nn \\
t_{nm}^{BC} & = & Tr(\si_n^{B}\otimes\si_{m}^{C}\rho_{BC}) = Tr(I^{A}\otimes\si_n^{B}  \otimes\si_{m}^{C}\rho),  \nn
\end{eqnarray}
\begin{equation}
\tau_{lmn}=Tr(\si_l^{A}\otimes\si_n^{B}\otimes\si_m^{C}\rho), \nn
\end{equation}
where $\rho_{A,B,C}$ and $\rho_{AB,AC,BC}$ are the appropriate reduced density operators. We consider the general interaction between qubits which can be reduced by the singular value decomposition to the Hamiltonian
\begin{equation}
H_{I}=H_{AB}+H_{AC}+H_{BC}   \label{eqn19}
\end{equation}
where
\begin{eqnarray}
H_{AB} & = & \sum_{s=1}^{3}\mu_s^{AB}\si_s^{A}\otimes\si_{s}^{B}\otimes I^{C} \nn \\
H_{BC} & = & \sum_{s=1}^{3}\mu_s^{BC}I^{A}\otimes\si_s^{B}  \otimes\si_{s}^{C} \nn \\
H_{AC} & = & \sum_{s=1}^{3}\mu_s^{AC}\si_s^{A}\otimes I^{B} \otimes\si_{s}^{C}.  \label{eqn20}
\end{eqnarray}
It is helpful to imagine that the three spins are at the vortices of a triangle. If they are arranged on a line, we expect on physical grounds that one of the terms can be neglected in comparison with the other two, as it gives the next nearest neighbor interaction. For all subsystems we have $\mu_1\geq\mu_2\geq\mu_3.$
Using the definition of the entanglement generation rate $\G$ in Eq. (3) and the definition of the entanglement measure in Eq. (7) we get,
\begin{equation}
\G=\frac{1}{||\mathcal{T}||}\sum_{i,j,k=1}^{3}\tau_{ijk}\dot{\tau}_{ijk}. \nn
\end{equation}
Using the Heisenberg equation of motion,
\begin{equation}
i\frac{d\rho}{dt}=[H_{I},\rho], \nn
\end{equation}
where the Hamiltonian $H_{I}$ is defined via Eq.s (\ref{eqn19}),(\ref{eqn20}), we get
\begin{equation}
\dot{\tau}_{ijk}=iTr(H_{I}[\si_i\otimes\si_j\otimes\si_k, \rho]). \nn
\end{equation}
We now use the commutator identity
$$[A\otimes B\otimes C, D\otimes E\otimes F]= \frac{1}{4}([A,D]\otimes[B,E]\otimes[C,F]+[A,D]\otimes\{B,E\}\otimes\{C,F\}$$
 $$+\{A,D\}\otimes[B,E]\otimes\{C,F\}+\{A,D\}\otimes\{B,E\}\otimes[C,F])$$
and the definitions of $\rho$ and $H_{I}$ in Eq.s (\ref{eqn15}) and (\ref{eqn19}) respectively to get,
{\setlength\arraycolsep{2pt}
\begin{eqnarray}
\dot{\tau}_{ijk}=-2\Big[\mu_k^{AC}\sum_{j^{\prime}=1}^{3}t_{j^{\prime}j}^{AB}\varepsilon_{ij^{\prime}k}+\mu_k^{BC}\sum_{k^{\prime}=1}^{3}t_{ik^{\prime}}^{AB}\varepsilon_{jk^{\prime}k}+ {} \nn \\
+\mu_j^{BC}\sum_{l^{\prime}=1}^{3}t_{il^{\prime}}^{AC}\varepsilon_{kl^{\prime}j}+
\mu_i^{AC}\sum_{l^{\prime}=1}^{3}t_{jl^{\prime}}^{BC}\varepsilon_{kl^{\prime}i}+ {} \nn \\
+\mu_i^{AB}\sum_{k^{\prime}=1}^{3}t_{k^{\prime}k}^{BC}\varepsilon_{jk^{\prime}i}
+\mu_j^{AB}\sum_{j^{\prime}=1}^{3}t_{j^{\prime}k}^{AC}\varepsilon_{ij^{\prime}j}\Big],  \label{eqn25}
\end{eqnarray}}
where $\varepsilon$ s are the Levi-Civita symbols. Substitution of Eq. (\ref{eqn25}) in the expression for $\G$ gives,
{\setlength\arraycolsep{2pt}
\begin{eqnarray}
\G=\frac{-2}{||\mathcal{T}||}\Big[\sum_{k,s=1}^{3}\left[(\vec{\tau}_{:sk}\times\vec{t}_{:k}^{AC})_s + (\vec{\tau}_{s:k}\times\vec{t}_{:k}^{BC})_s\right]\mu_s^{AB} + {} \nn \\
+\sum_{i,s=1}^{3}\left[(\vec{\tau}_{i:s}\times\vec{t}_{i:}^{AB})_s + (\vec{\tau}_{is:}\times\vec{t}_{i:}^{AC})_s\right]\mu_s^{BC} + {} \nn \\   +\sum_{j,s=1}^{3}\left[(\vec{\tau}_{:js}\times\vec{t}_{:j}^{AB})_s + 
(\vec{\tau}_{sj:}\times\vec{t}_{j:}^{BC})_s\right]\mu_s^{AC}\Big],  \label{eqn26}
\end{eqnarray}}
where $\vec{\tau}_{:sk}=[\tau_{1sk},\tau_{2sk},\tau_{3sk}]^{T},$ for example, is a vector in $\mathbb{R}^3$ for fixed $s$ and $k.$ Similarly, 
$\vec{t}_{:k}^{\cdot\cdot}$ and $\vec{t}_{j:}^{\cdot\cdot}$ are the $k$th column and the $j$th row vectors of the matrix $t^{(\cdot\cdot)}.$  
The expression for the entanglement generation rate $\G$ for three qubits (Eq. (\ref{eqn26})) has 54 coupled terms, each term being a component of the cross product of two vectors. A geometric argument to maximize $\G ,$ as in the two qubit case, seems to be very difficult. However, it is quite straightforward to maximize $\G$ numerically, by writing the elements of the three way array $\mathcal{T}$ and the matrices $t^{(\cdot\cdot)}$ as the appropriate averages in the general three qubit state

\begin{equation}
|\psi\ran = \sum_{i=0}^{7}c_i|i\ran\;\;\;;\;\;\;\sum_i|c_i|^2=1  \label{eqn27}
\end{equation}
where $i$ labeling the product basis ket $|i\ran$ is the binary representation of the index $i.$ 
We can numerically optimize $\G$ for the general Hamiltonian in Eq. (\ref{eqn19}). However, for the general case, where the interaction is anisotropic, that is, the strengths of interaction $\mu_k^{(\cdot\cdot)}$ have different values, $\G$ does not have the simple structure $\G=f(p)h_{max}$ as in the two qubit case. Therefore, we assume isotropic interactions so that all interaction strengths are equal to a common value $\mu .$ In this case, after evaluating all the terms in Eq. (\ref{eqn26}) in the state $|\psi\ran$ given by Eq. (\ref{eqn27}), $\G$ can be written as
\begin{equation}
\G= h(c_0,\ldots,c_7)\mu. \nn
\end{equation} 
After the numerical optimization of $\G$ as a function of $c_i, \; i=0,1,\ldots,7,$ we get,
\begin{equation}
\G^{max}= 5.72523 \mu.  \nn 
\end{equation}        
The (normalized) state corresponding to this $\G^{max}$ is given by
$$|\psi_{max}\ran= (0.033768-i0.168758|000)\ran+(0.574022-i0.0709471)|001\ran+$$ $$+(0.0218412-i0.111565)|010\ran+  
 (0.672021-i0.0754116)|011\ran+$$ $$+(-0.0603488+i0.172566)|100\ran
+(-0.0051137-i0.183831)|101\ran+ $$
$$+(0.0556843+i0.151888)|110\ran+(0.0700719-i0.259423)|111\ran .$$
This state has the following Acin canonical form, expressed by the two fold degenerate sets of entanglement parameters [22].
$$|\psi^+\ran=0.610291|000\ran+0.67402\exp(i2.51395)|100\ran+0.394893|101\ran +$$
$$+0.110357|110\ran+0.0715772|111\ran ,$$
or,
$$|\psi^-\ran=0.329873|000\ran+0.546087\exp(i0.402558)|100\ran+0.730583|101\ran+$$
$$+0.20417|110\ran+0.132424|111\ran .$$
We see that the state with the maximal entanglement generation rate $\G^{max}$ belongs to the GHZ class. Further, we find that $E(|\psi_{max}\ran)=0.258918.$ This means that, given the isotropic interaction, it is beneficial to start with an entangled three qubit state for optimal entanglement generation. Also, we note that the optimal entanglement is independent of $H_{I},$ provided the corresponding interaction is isotropic. 

Thus we see that, for three qubits, the geometric method based on the entanglement measure given by Eq. (7) can be numerically implemented to get the state with maximal entanglement generation rate. This program can be carried out for the general interaction Hamiltonian in Eq. (\ref{eqn19}) although we have restricted to the isotropic interactions. This procedure can be suitably carried out in a laboratory using quantum circuits. Every quantum circuit acts unitarily on a quantum state and we can always find a Hamiltonian corresponding to such a circuit [9, 23]. On the other hand, given a (interaction) Hamiltonian  for a three qubit system, we may construct a circuit implementing the corresponding evolution using universal quantum gates. We note that, for the isotropic interaction, the maximal entanglement generation rate $\G^{max}$ is proportional to the interaction strength $\mu$ and the corresponding state is independent of $\mu,$ as in the two qubit case. When the interactions are anisotropic, the scenario for two qubits does not apply to the three qubit case, as $\G$ does not factor into the product of a state dependent function and an expression involving only the interaction strengths. Even when the interactions are isotropic, we do not know the explicit form of such a state dependent function. In other words, we do not know whether it is possible to separately account for the contribution due to the state and that due to the interactions. Thus a general procedure for the maximization of the entanglement generation rate for the higher dimensional and multipartite systems still seems to be an open question. These observations ensue from the fact that the terms in the expression for $\G$ could not be decoupled. This difficulty seems to be generic, as it may be a consequence of the difficulties in the geometric interpretation of the Bloch space for the multipartite and higher dimensional systems [24].  All the  remarks in this paragraph apply to the two qutrit case as well. \\

\textbf{ACKNOWLEDGMENTS}\\
 
 We thank A. K. Pati for suggesting this problem and pointing out reference [3]. We thank Guruprasad Kar and Prof. R. Simon for encouragement. We thank Sibasish Ghosh for a useful discussion. We thank P. Durganandini for a useful discussion. ASMH thanks Sana'a university for financial support.\\

\textbf{REFERENCES} \\

\begin{verse}

[1] Dirk Bouwmeester, A. Ekert, A. Zeilinger (Eds.), \textit{The Physics of Quantum Information}, ( Springer-Verlag Berlin Heidelberg 2000); C. H. Bennett, G. Brassard, C. Crépeau, R. Jozsa, A. Peres, and W. K. Wootters, Phys. Rev. Lett. \textbf{70}, 1895 (1993).

[2] Ye Yeo and Wee Kang Chua, Phys. Rev. Lett. 96, 060502 (2006).

[3] W. Dur, G. Vidal, J. I. Cirac, N. Linden, and S. Popescu, Phys. Rev. Lett. \textbf{87}, 137901 (2001).

[4] C. H. Bennett, A. Harrow, D. W. Leung, and J. A. Smolin, IEEE Trans. Inf. Theory, Vol. \textbf{49}, p.1895-1911  (2003),  quant-ph/0205057.

[5] P. Zanardi, C. Zalka, and L. Faoro, Phys. Rev. A \textbf{62}, 030301(R) (2000); B. Kraus and J. I. Cirac, Phys. Rev.
A \textbf{63}, 062309 (2001); M. S. Leifer, L. Henderson, and N. Linden, Phys. Rev. A \textbf{67}, 012306 (2003).

[6] D. Beckman, D. Gottesman, M. A. Nielsen, and J. Preskill, Phys. Rev. A \textbf{64}, 052309 (2001).

[7] K. Hammerer, G. Vidal, and J. I. Cirac, Phys. Rev. A \textbf{66}, 062321 (2002).

[8] D. W. Berry and B. C. Sanders, Phys. Rev. A. \textbf{67}, 040302(R) (2003).

[9] J. L. Dodd, M. A. Nielsen, M. J. Bremner, and R. T. Thew, Phys. Rev. A \textbf{65}, 040301(R) (2002); P. Wocjan,
D. Janzing, and Th. Beth, Quantum Information and Computation \textbf{2}, 117 (2002); N. Khaneja, R. Brockett, and S. J. Glaser, Phys. Rev. A \textbf{63}, 032308 (2001); G. Vidal and J. Cirac, Phys. Rev. A \textbf{66}, 022315 (2002); P.
Wocjan, M. Roetteler, D. Janzing, and Th. Beth, Quantum Information and Computation \textbf{2}, 133 (2002), Phys.
Rev. A \textbf{65}, 042309 (2002); M. A. Nielsen, M. J. Bremner, J. L. Dodd, A. M. Childs, and C. M. Dawson, Phys. Rev.
A \textbf{66}, 022317 (2002); H. Chen, quant-ph/0109115; G. Vidal, K. Hammerer, and J. I. Cirac, Phys. Rev. Lett. \textbf{88}, 237902 (2002); Ll. Masanes, G. Vidal, and J. I. Latorre, Quantum Information and Computation \textbf{2}, 285 (2002).

[10] C. H. Bennett, J. I. Cirac, M. S. Leifer, D. W. Leung, N. Linden, S. Popescu, and G. Vidal, Phys. Rev. A \textbf{66}, 012305 (2002).

[11] G. Vidal and J. I. Cirac, Phys. Rev. Lett. \textbf{88}, 167903 (2002).

[12] Special issue, Fortschr. Phys. \textbf{48} No. 9-11 (2000).

[13] C. H. Bennett and S. J. Wiesner, Phys. Rev. Lett. \textbf{69}, 2881 (1992).

[14] C. H. Bennett, G. Brassard, C. Cr´epeau, R. Jozsa, A. Peres, and W. K. Wootters, Phys. Rev. Lett. \textbf{70}, 1895
(1993).

[15] Ali Saif M. Hassan and Pramod S. Joag, Phys. Rev. A \textbf{77}, 062334 (2008).

[16] Ali Saif M. Hassan and Pramod S. Joag,  Quantum Information and Computation \textbf{8}, 0773 (2008).

[17] G. Mahler and V.A. Weberru\ss, {\it Quantum Networks}, (Springer , Berlin, 1995).

[18] Julio I. de Vicente,  J. Phys. A: Math. Theor. \textbf{41}, 065309 (2008)

[19] J. Eisert, K. Jacobs, P. Papadopoulos, and M. B. Plenio, Phys. Rev. A \textbf{62}, 052317 (2000).

[20] Asher Peres, {\it Quantum Theory: Concepts and Methods}, (Kluwer Academic Publishers, 1993).

[21] Ali Saif M. Hassan and Pramod S. Joag, Phys. Rev. A \textbf{80}, 042302 (2009). 

[22] A Acin, A Andrianov, E Jane and R Tarrach, J. Phys. A \textbf{34}, 6725-6739 (2001).

[23] B.Kraus and J.I.Cirac Phys. Rev. A \textbf{63}, 062309 (2001).

[24] G.Kimura and A. Kossakowski Open Systems and Information Dynamics \textbf{12}, 207-229 (2005).

\end{verse}
\end{document}